\documentclass[twocolumn,aps,prl,showpacs,floatfix]{revtex4}
\usepackage{amsmath,bm}
\usepackage{graphicx}
\usepackage{epsfig}
\begin{document}

\title{Shear viscosity to entropy density ratio of a relativistic Hagedorn resonance gas}
\author{Subrata Pal}
\affiliation{Department of Nuclear and Atomic Physics, Tata Institute of
Fundamental Research, Homi Bhabha Road, Mumbai 400005, India} 

\begin{abstract} 
The new state of matter produced at Relativistic Heavy Ion Collider reveals a strongly 
coupled quark-gluon plasma with an extremely small shear viscosity to entropy density 
ratio $\eta/s$. We calculate the $\eta/s$ of an equilibrated hadron matter characterized 
by a relativistic hadron resonance gas with a Hagedorn mass spectrum that grows 
exponentially with the hadron mass. We find with increase in temperature of the system 
the $\eta/s$ value decreases due to rapid increase in the multiplicity of massive 
resonances. In the vicinity of the critical temperature for deconfinement transition, 
the minimum value of $\eta/s$ in the Hagedorn resonance gas is found to be 
consistent with the current estimates for a strongly coupled quark-gluon plasma.
\end{abstract}

\maketitle

Heavy ion collisions at the BNL Relativistic Heavy Ion Collider (RHIC) have revealed a 
new state of matter \cite{BRAHMS,PHOBOS,STAR,PHENIX} comprising of strongly interacting 
quarks and gluons (sQGP) \cite{DIS}. This conclusion is based on viscous hydrodynamic model 
analysis of elliptic flow that requires a very small shear viscosity to entropy density 
ratio $\eta/s$ of $0.08 - 0.24$ \cite{Mike,Paul,Song}. The main uncertainty in these 
estimates stems from the equation of state and the initial conditions employed. 
This value is remarkably similar to the lower bound $\eta/s \geq 1/4\pi$ obtained by 
Kovtun-Son-Starinets (KSS) \cite{KSS} for infinitely coupled supersymmetric Yang-Mills gauge 
theory based on the AdS/CFT duality conjecture. A recent lattice calculation \cite{Meyer} of 
gluonic plasma do support the current estimates. While leading order perturbative 
calculations result in a significantly large value of $\eta/s \approx 0.8$ for 
$\alpha_s =  0.3$ \cite{Arnold}.

It has been also argued \cite{Csernai,Lacey} that with increasing temperature in the 
hadronic phase $\eta/s$ decreases and reaches a minimum at or near the critical temperature 
$T_c$ and increases thereafter in the deconfined phase. Indeed this behavior has 
been observed in several substances in nature all of which satisfy the KSS bound 
suggesting the bound could be universal. However, no consensus have yet been reached 
\cite{Blaizot} on the physical mechanism that accounts for the thermodynamic and transport 
properties of the system that leads to the minimal viscosity. Understanding the transport 
coefficients of the matter formed at RHIC is thus very challenging. Since in a heavy ion 
collision the system evolves from a QGP phase to the confined hadrons (low temperature 
phase of QCD), it is important to understand systematically the transport properties 
of a hadronic system in order to ascertain the sQGP properties with minimal uncertainty.
In this letter, we investigate the transport properties of a infinite equilibrated 
hadronic matter comprising of massive hadronic resonances $-$ the Hagedorn states 
\cite{Hagedorn,Kapusta,Venu} and all the low-lying observed hadrons within a Monte 
Carlo sequential binary emission model \cite{PalPD,Pal}. 

Several calculations of $\eta/s$ in the hadron phase have been performed with various 
techniques such as the chiral perturbation theory, linearized Boltzmann equation
\cite{Dobado,Chen,Itakura}, and microscopic transport theory \cite{Muronga}. 
However, most of these calculations employ at most two component system or mesonic gas 
with a wide variation in the estimate of $\eta/s \sim 0.08-1 $ at $\sim T_c$.  
A recent microscopic transport calculation \cite{Bass} within the UrQMD model accounted 
all the measured hadron species of mass $m \lesssim 2$ GeV yields a minimum 
$\eta/s \approx 0.9$ at zero baryon chemical potential. This is significantly higher 
than in the viscous hydrodynamic estimate. Clearly it indicates the importance of massive 
resonances on the transport properties of hadron gas in the strong interaction domain of 
QCD matter especially near the critical temperature $T_c = 196$ MeV predicted in 
lattice simulations \cite{Cheng}.  

On the other hand, it was proposed \cite{Hagedorn} that the density of hadronic 
states grows exponentially with resonance mass $m$, 
$\rho_{\rm HS}(m) \sim m^{-\alpha} \exp(m/T_H)$, where $T_H \sim 150-200$ MeV is
the Hagedorn temperature. The measured hadronic states up to $m \sim 2$ GeV are indeed
qualitatively consistent with the Hagedorn mass spectrum \cite{Bron}. 
In absence of hadronic interaction, the energy density for
such a system will diverge leading to a maximum/limiting temperature $T_H$ for the 
hadronic matter. In the infinite mass limit, the thermodynamical quantities for the 
Hagedorn resonance gas would thus show critical behavior as it crosses $T_H \approx T_c$
which may be associated with deconfining transition \cite{Castorina}. It has been 
also argued that the idea of Hagedorn states (HS) arise naturally in QCD at large $N_c$ 
\cite{Cohen} and their indirect evidence can be found in lattice studies of thermodynamic 
properties of hadron matter above the deconfinement temperature \cite{Bringoltz}. 
For a hadron resonance gas with Hagedorn states the reduction of $\eta/s$ in the 
vicinity of $T_c$ has been demonstrated \cite{Noronha} within the rate equation approach.

\begin{figure}[ht]
\centerline{\epsfig{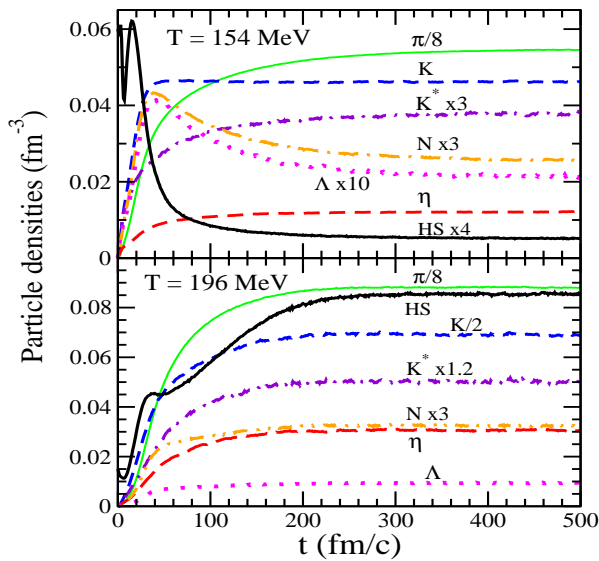}}
\vspace{-0.2cm}

\caption{Time evolution of particle densities for an equilibrated system of hadron 
resonance gas with Hagedorn states at a zero net baryon density at temperatures
$T = 154$ (top panel) and 196 (bottom panel) MeV.}

\label{mulp}
\end{figure}

In the present study of the properties of an equilibrated Hagedorn gas in the 
binary emission model \cite{PalPD,Pal}, we include the Hagedorn states 
(mass $\gtrsim 2$ GeV) and all the low-lying measured hadrons available in the 
Particle Data Book. Detail description of the decay widths of HS and
other resonances and their formation cross section in binary collisions during dynamic 
evolution of the system can be found in Ref. \cite{PalPD,Pal}. For brevity, we mention that 
depending on the mass, a HS may undergo binary decay into one of the three possible channels 
(i) two observed discrete hadrons (DH) of $m < 2$ GeV, (ii) a DH and a HS, (iii) two HSs. 
Based on Hagedorn hypothesis the density of massive states are assumed to be
\begin{equation} \label{dos}
\rho(m,q) = A\frac{\exp\!\left[\left\{m-m_gf(m-m_g)\right\}/T_H\right]}
{\left[\left\{m-m_gf(m-m_g)\right\}^2 + m_r^2\right]^\alpha} ,
\end{equation}
in the usual notation \cite{PalPD,Pal}. Here $m_r = 0.5$ GeV, and a HS of mass $m$ is 
characterized by its baryon, strangeness, spin and isospin quantum numbers 
$q=(B,S,J,I,I_z)$ with a ground state mass $m_g = a_Q({\rm max}|3B+S|,2I) + a_S|S|$; 
the parameters ($a_Q,a_S$) are determined empirically 
from the measured smallest masses. This model was found to successfully describe the 
stable and resonance yield ratios and their spectra at RHIC. We consider here the Hagedorn
temperature $T_H = 196$ MeV consistent with the critical temperature $T_c$ in lattice QCD 
prediction of a crossover transition at vanishing baryon chemical potential \cite{Cheng}. 
A resulting exponent $\alpha = 2.44$ is then estimated by comparing the theoretical and 
experimental cumulants of the spectrum \cite{Bron,PalPD}.

To generate the equilibrated matter we initiate a system of Hagedorn states in a box 
with periodic boundary conditions in configuration space. Subsequent binary decay and 
their regeneration in binary collision during time evolution enforce the system to 
thermodynamic equilibrium. Chemical equilibrium is determined from the
saturation of various particle densities when their production and annihilation rates 
become identical. While kinetic equilibration occurs when the system approaches momentum 
isotropization. By fitting the particle energy spectra with a Boltzmann distribution 
$dN_i/d^3p \sim \exp(-E_i/T^*)$, thermal equilibration is verified when all the particle 
species at later times can be described by the same slope temperature $T^*$.
For point-particles employed in our simulation, the pressure can be evaluated from the 
virial theorem as 
\begin{equation} \label{pres}
P_{\rm pt}(T^*) = \frac{1}{3V} \sum_{i=1}^{N_{\rm part}} \frac{|{\bf p}|^2_i}{p^0_i} ~,
\end{equation}
where ${\bf p}_i$ and $p^0_i$ are the momenta and energy of the $i$th particle. 
While the inclusion of Hagedorn states in the system provides an attractive interaction,
use of point particles in the simulation ignores the repulsive interactions among
the finite size hadrons especially the massive HS \cite{Hagedorn,Kapusta,Venu}
that would drive the system to deconfinement at $T_c$. The effects of repulsive interaction 
can be included via excluded-volume approach \cite{Kapusta} where the volume occupied 
by $i$th hadron in the relativistic approximation is $V_i = p^0_i/4B$, where 
$B$ is the effective MIT bag constant. The excluded-volume temperature and pressure
are related to their point-particle analogs as \cite{Kapusta}
\begin{equation} \label{expres}
T = \frac{T^*}{1-P_{\rm pt}(T^*)/4B}, ~~~
P(T) = \frac{P_{\rm pt}(T^*)}{1-P_{\rm pt}(T^*)/4B} ~.
\end{equation}
The other thermodynamic quantities can be obtained using the thermodynamic identities.
We consider $B^{1/4} = 0.40$ GeV that results in $P(T_c) < 4B$ with a corresponding
limiting temperature $T_{\rm lim} > T_c$ \cite{Kapusta} as evident from Eq. (\ref{expres}). 
We restrict our calculations to $T \leq T_c$ in the confined hadronic system.

\begin{figure}[ht]
\centerline{\epsfig{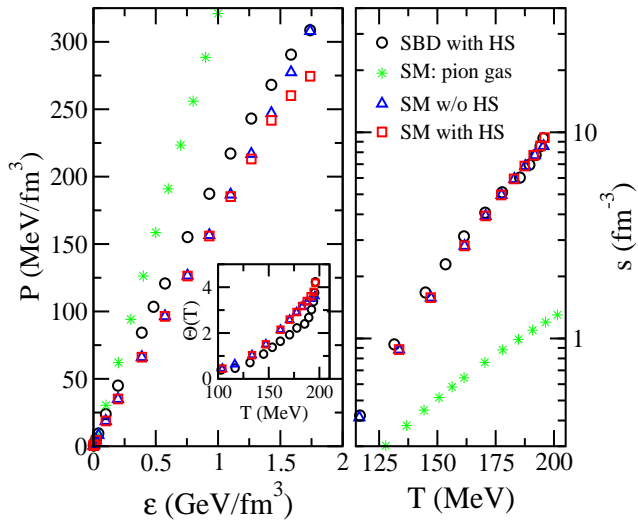}}
\vspace{-0.2cm}

\caption{Pressure versus energy density (left panel), entropy density versus temperature 
(right panel), and the interaction measure $\Theta = (\varepsilon-3P)/T^4$ (inset). 
The results are in sequential binary decay (SBD) model for hadron gas with Hagedorn
states (HS) (circles) and in the statistical model (SM) for an pion gas (stars), 
a hadron resonance gas without HS (triangles), and with inclusion of HS (squares).}

\label{eos}
\end{figure}

Figure \ref{mulp} shows the time evolution of various particle abundances at zero net 
baryon $\rho_B$ and strangeness $\rho_S$ densities. At a temperature $T \approx 154$ MeV, 
the initial distribution of massive Hagedorn states ($m \sim 20$ GeV) decay and 
regenerate within few fm/c. Subsequently the HS multiplicity 
decreases rapidly to a small value with simultaneous production of lighter ($m < 2$ GeV) 
hadrons. In contrast, at $T \approx 196$ MeV each massive HS decays dominantly to a 
lighter HS and a discrete hadron resulting in a continuous increase in the density of 
Hagedorn states and other hadron resonances. At about $t = 200$ fm/c the particle 
densities saturate suggesting chemical equilibration has been achieved. 

In Fig. \ref{eos} (left panel) we present the equation of state, i.e. pressure versus 
energy density, of an equilibrated hadron matter at zero net baryon density. With
increasing energy density when the temperature $T \sim T_c$, enhanced particle production 
of especially heavy resonances and HS causes softening of the equation of state. 
Consequently the speed of sound $c^2_s = (\partial P/\partial\varepsilon)_V$ in the medium 
drops gradually from 
$c^2_s \approx 0.18$ at $T \approx 100$ MeV to about 0.09 at $T_c$. The estimated minimum 
speed of sound is consistent with the lattice data \cite{Cheng}. In contrast, the UrQMD 
model hadron resonance gas (without HS) yields a constant value of $c^2_s = 0.18$ \cite{Bass}.

To gauge the dynamics of hadron-hadron interaction, we also investigate the thermodynamic 
properties in an independent statistical model for an hadron resonance gas in the 
grand canonical ensemble. The particle densities in the statistical model is given by 
\cite{Belkacem}
\begin{equation} \label{SM}
\rho_i = \frac{g_i}{2\pi^3} \int \rho_i(m) \: dm \int_0^\infty 
\frac{4\pi p^2 dp}{\exp(p^0_i -\mu_i)/T^* \pm 1} ~,
\end{equation}
where the $\pm$ sign refers to fermions/bosons. For the $i$th species, $g_i$ is the spin-isospin
degeneracy and $\mu_i = \mu_B B_i + \mu_S S_i$ is the chemical potential with $\mu_B$ and $\mu_S$
the baryon and strangeness chemical potentials, respectively. For the observed resonances, 
$\rho_i(m)$ is taken as Breit-Wigner mass distribution while for HS it is replaced by 
$\rho_{\rm HS}$ of Eq. (\ref{dos}). The results are found to be rather insensitive to the choice
of upper mass limit of integration for HS $m_{\rm max} \geq 40$ GeV. In the limit 
$m_{\rm max} \to \infty$, as the estimated exponent $\alpha = 2.44$ in Eq. (\ref{dos}), the 
point-particle energy density, pressure and entropy density of an ideal classical (Boltzmann) 
Hagedorn gas remain finite at $T \leq T_H$ and diverge at $T > T_H$ exhibiting critical behavior 
\cite{Castorina}; albeit in the excluded-volume approach, the corresponding quantities evaluated 
at $T \leq T_H=T_c$ would remain finite.

In Fig. \ref{eos} (left panel) we also show the equation of state obtained in the statistical 
model. Compared to a pion gas (stars), inclusion of measured discrete resonances 
up to mass $\sim 2$ GeV (triangles) results in a considerable decrease in the system pressure. 
The interaction measure, $\Theta = (\varepsilon - 3P)/T^4$, then gradually increases
as $T \to T_c$. Further inclusion of Hagedorn states in the statistical model opens up 
additional degrees of freedom and provides an attractive interaction that slows the pressure 
increase at $\varepsilon \gtrsim 1.5$ GeV/fm$^3$. The trace anomaly $\Theta$ then increases 
dramatically near $T = T_H$. The speed of sound in the statistical model with HS turns out to be
rather small $c^2_s \approx 0.07$ at $T_c$. Comparison of this curve with sequential binary 
emission model result provides a measure of the dynamics of repulsive interaction in the medium. 
The interaction effect starts at temperature above the pion mass and becomes significant at 
high energies where massive resonances are produced. 

The point-particle entropy density of the system can be computed using the Gibbs formula
\begin{eqnarray} \label{entr}
s_{\rm pt}(T^*) &=& (\varepsilon_{\rm pt} + P_{\rm pt} 
- \sum_i\mu_i\rho_i)/T^* \nonumber\\
&=& (\varepsilon_{\rm pt} + P_{\rm pt} - \mu_B\rho_B)/T^*  ~,
\end{eqnarray}
where $\rho_B = \sum_i B_i \rho_i$ is the net baryon density and the (initial) net 
strangeness density $\rho_S$ is set to zero in our calculation. The baryon chemical 
potential $\mu_B$ is evaluated from the particle yield ratios at equilibrium. 
Figure \ref{eos} (right panel) shows the excluded-volume entropy density \cite{Kapusta}
as a function of temperature at zero baryon chemical potential. In the statistical model, 
compared to the pion gas, larger degrees of freedom in the hadron (and Hagedorn) 
resonance gas at the same energy density lowers its temperature which results in higher 
entropy \cite{PalPratt}. In the sequential emission model the increase (or slope) of 
entropy density with temperature near $T_c$ is somewhat less dramatic and captures 
the trend associated with that of a crossover transition obtained in the lattice 
data \cite{Cheng}. 

We now extract the shear viscosity for the dynamically evolving system of an equilibrated 
hadron resonance gas with Hagedorn states using the Kubo relation \cite{Kubo}. The
transport coefficients determine the dynamics of fluctuations of dissipative fluxes 
about the equilibrated state. The Kubo formalism employs the linear response theory
to relate the transport coefficients as correlations of dissipative fluxes by considering 
the dissipative fluxes as perturbations to local thermodynamic equilibrium 
\cite{Kubo,Hosoya,Paech}. The Green-Kubo relation for shear viscosity is
\begin{equation} \label{visco}
\eta  = \frac{1}{T} \int d^3r \int_0^\infty dt \: \langle \pi^{xy}({\bf 0}, 0)
\pi^{xy}({\bf r}, t) \rangle_{\rm eqlb} ~ .
\end{equation}
Here $t$ refers to time after the system equilibrates which is set at $t = 0$, and $\pi^{xy}$ 
is the shear component (traceless part) of the energy momentum tensor $\pi^{\mu\nu}$:
\begin{equation} \label{shear}
\pi^{xy} = \int d^3p \: \frac{p^x p^y}{p^0} f(x,p)  \: \equiv \:
\frac{1}{V} \sum_{i=1}^{N_{\rm part}} \frac{p^x_i p^y_i}{p^0_i} ~.
\end{equation}
The equivalence accounts for the point particles used in our simulation. The averaging 
in Eq. (\ref{visco}) is over the ensemble of events generated in the simulation.

\begin{figure}[ht]
\centerline{\epsfig{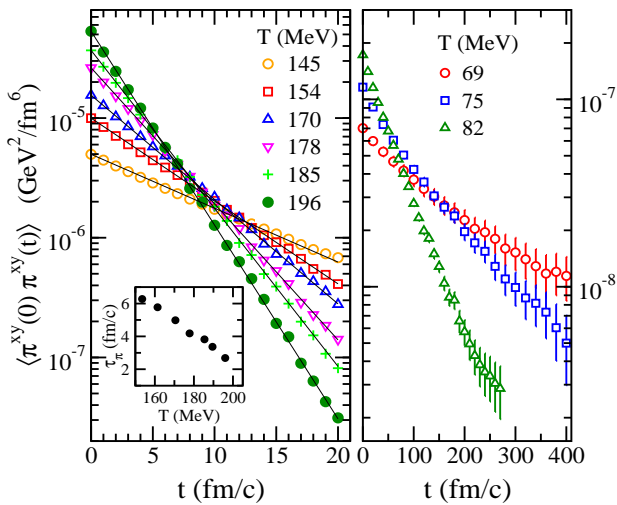}}
\vspace{-0.2cm}

\caption{Shear-stress correlation for the hadron resonance gas with Hagedorn 
states as a function of time at different system temperatures.
The inset shows the temperature versus relaxation time $\tau_\pi$ obtained 
from fits to the correlations using an exponential decay.}

\label{corr}
\end{figure}

Figure \ref{corr} displays the correlation function of the shear-stress tensor as a 
function of time. The correlation functions are found to damp exponentially with time, 
i.e. $\langle \pi^{xy}(0)\pi^{xy}(t)\rangle \sim \exp(-t/\tau_\pi)$. Note at small 
temperatures (right panel) the correlations sustain for long times however show a 
power law tail. The relaxation times $\tau_\pi$ for the shear fluxes so obtained by 
fitting the correlation functions are found to decrease with increasing temperature. 
The shear viscosity is estimated from Eq. (\ref{visco}) by integrating the correlation 
function over time using the exponential decay found empirically.

\begin{figure}[ht]
\centerline{\epsfig{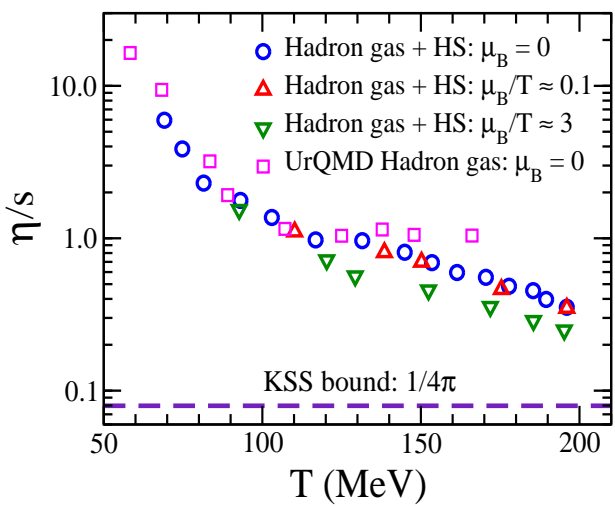}}
\vspace{-0.2cm}

\caption{The shear viscosity to entropy density ratio $\eta/s$ as a function of temperature 
for a hadron resonance gas with Hagedorn states in thermodynamic equilibrium at various
baryon chemical potential $\mu_B$. The $\eta/s$ for a pure hadron resonance gas in the 
UrQMD model \cite{Bass} is also shown (squares).}

\label{etas}
\end{figure}

The shear viscosity to entropy density ratio $\eta/s$ for the Hagedorn resonance gas in 
equilibrium at zero baryon chemical potential is presented in Fig. \ref{etas}. Also shown
in the figure are the $\eta/s$ values for hadron resonance gas without Hagedorn states in 
the UrQMD model \cite{Bass}. Our results for $\eta/s$ are comparable to the UrQMD estimates
at temperatures $T < 150$ MeV. At higher temperatures, massive resonances contribute 
dominantly resulting in gradual reduction of $\eta/s$ with temperature \cite{Noronha} 
and reaches a minimum of $(\eta/s)_{\rm min} \approx 0.34$ at $T_c = 196$ MeV.
In contrast, the UrQMD model calculation 
shows almost a constant value for $\eta/s \approx 0.9$ at $T \approx 125-166$ MeV. 
The reduction of $\eta/s$ with Hagedorn mass spectrum can be realized as primarily due 
to enhancement of massive resonances that leads to a decrease in the mean free path 
of a particle and a corresponding increase in the average binary collision cross section 
$\sigma$. In classical transport theory, as $\eta \sim \overline p/\sigma$ 
(where $\overline p$ is the average momentum of the particle), implies a reduction 
of shear viscosity. The $(\eta/s)_{\rm min} \approx 0.34$ at zero baryon chemical potential 
for the equilibrated Hagedorn resonance gas is significantly above the KSS bound of 
$1/4\pi$ \cite{KSS} while somewhat close to the upper bound of 0.24 obtained from viscous 
hydrodynamic analysis \cite{Paul,Song} of elliptic flow; albeit $\eta/s$ is assumed constant 
throughout the hydrodynamic evolution of the system. Note however, for Hagedorn resonance 
gas treated within the rate equation approach \cite{Noronha}, an upper limit of $\eta/s$ 
estimated in the kinetic theory is found to be as small as the KSS bound near $T_c$.

The measured antibaryon-to-baryon ratio $\overline B/B \approx 0.8$ at midrapidity for 
central Au+Au collision at RHIC \cite{BRAHMS,PHOBOS,STAR,PHENIX} suggests the hadronic matter 
formed possess, though small, but a finite net baryon number density $\rho_B$. During 
the late hadronic stage of evolution essentially from chemical equilibrium to thermal 
equilibrium the system could acquire non-equilibrium chemical composition \cite{Hirano}. 
In fact, reproduction of the measured particle yield ratios at RHIC requires a non-zero baryon 
chemical potential in the sequential binary emission model \cite{PalPD,Pal}, while
thermal model fits \cite{Braun,Rafelski} to the Au+Au collision data at the RHIC 
energy of $\sqrt s = 200$ GeV yields a chemical freeze-out temperature of 
$T \sim 160$ MeV and baryon chemical potential $\mu_B \sim 20$ MeV.
In Fig. \ref{etas} we show $\eta/s$ as a function of 
temperature at a finite baryon chemical potential ($\mu_B/T \approx 0.13$) created by 
inducing a finite $\rho_B$ in the initial distribution of Hagedorn states. At a
given $T$, the energy density of the system is larger for non-zero baryon chemical potential 
leading to further enhancement of particularly massive particle abundances. 
In the $T-\mu_B$ region explored, the shear viscosity $\eta$ {\em increases}
compared to that at $\mu_B = 0$. 
This increment can be understood classically $-$ substantially large particle densities
though increase $\sigma_i$ and thereby reduces $\eta_i$ of each species $i$, however the 
total contribution, $\eta = \sum_i \eta_i$, from these species is greatly enhanced 
\cite{Itakura}. Since the entropy density of the system also grows with $T$, the resulting 
$\eta/s$ at $\mu_B/T \approx 0.13$ is found to be similar to $\mu_B =0$.
In Fig. \ref{etas} we also illustrate the effect of high $\mu_B/T \approx 3.0$ 
associated with heavy ion collisions at the top AGS energy of $\sqrt s = 4.85$ GeV 
and at the lowest CERN/SPS energy of $\sqrt s = 6.27$ GeV \cite{Braun,Rafelski}; 
albeit the temperatures reached in these reactions are much smaller 
at $T \sim 125-145$ MeV. We find that only for such large $\mu_B$ the extracted $\eta/s$ 
decreases further. The present study also suggests the need to incorporate, in the viscous 
hydrodynamics calculations, the Hagedorn states in the after-burning hadronic phase 
to quantify the dissipation and thus $\eta/s$ ratio. 

While weakly coupled perturbative QCD estimate of $\eta/s \approx 1$ at $T\gg T_c$, 
the QGP formed at RHIC in the region $T_c \leq T \lesssim 2T_c$ is thought to be strongly 
interacting. The current estimate of $\eta/s \sim 0.08-0.24$ \cite{Mike,Paul,Song,Meyer,PalAMPT}
at RHIC may be related to a minimum reached in the deconfined phase just above $T_c$. 
Our estimate of $\eta/s$ for the Hagedorn resonance gas suggests that, as the sQGP cools, 
the $\eta/s$ in deconfined phase makes a relatively continuous/smooth transition into 
the confined phase near $T_c$ without any discontinuity or a sharp increase proposed 
\cite{Bass} in absence of Hagedorn states.

In summary, we have studied the thermodynamic and transport properties of an infinite 
equilibrated matter composed of interacting hadron resonance gas plus an exponentially 
increasing Hagedorn mass spectrum. At temperatures close to the Hagedorn temperature, 
$T_H = T_c = 196$ MeV, the particle densities of the massive resonances
are enhanced significantly. This leads to a sharp rise in the trace anomaly and the 
entropy density of the system which captures the trend seen in the lattice data at
vanishing baryon chemical potential. The shear viscosity to entropy density ratio of 
the Hagedorn resonance gas, both in and out of chemical equilibrium, decreases with
temperature and baryon chemical potential and reaches a minimum of 
$(\eta/s)_{\rm min} \approx 0.3$ at about $T_c$. The extracted minimum value is 
near the upper bound of current estimates for a strongly coupled quark-gluon plasma 
formed in ultra-relativistic heavy ion collisions at RHIC.
\medskip

I would like to thank Steffen Bass, Rajeev Bhalerao and Pawel Danielewicz for many 
helpful discussions.

\end{document}